\documentclass{article}%
\usepackage{amsmath}
\usepackage{amsfonts}
\usepackage{amssymb}
\usepackage{graphicx}%
\providecommand{\U}[1]{\protect\rule{.1in}{.1in}}

\begin{document}

\title{Entropic Time\thanks{Presented at MaxEnt 2010, the 30th International Workshop
on Bayesian Inference and Maximum Entropy Methods in Science and Engineering
(July 4-9, 2010, Chamonix, France).} }
\author{Ariel Caticha\\{\small Department of Physics, University at Albany-SUNY, }\\{\small Albany, NY 12222, USA.}}
\date{}
\maketitle

\begin{abstract}
The formulation of quantum mechanics within the framework of entropic dynamics
includes several new elements. In this paper we concentrate on one of them:
the implications for the theory of time. Entropic time is introduced as a
book-keeping device to keep track of the accumulation of changes. One new
feature is that, unlike other concepts of time appearing in the so-called
fundamental laws of physics, entropic time incorporates a natural distinction
between past and future.

\end{abstract}

\section{Introduction}

A general framework for dynamics based on the method of maximum entropy is
applied to non-relativistic quantum mechanics. The basic assumption of
entropic dynamics is that in addition to the particles of interest there exist
other variables, which we may call \textquotedblleft hidden\textquotedblright,
to which we can associate an entropy. The evolution of the particles is a
diffusion driven by the entropy of the hidden variables. The second important
assumption is that there is a conserved energy. In \cite{Caticha
09}\cite{Caticha 10} it was shown that such a conservative diffusion is
equivalent to the Schr\"{o}dinger equation.

Entropic dynamics differs from other approaches to quantum mechanics in
several important aspects. One is the explicitly epistemological emphasis: the
laws of physics in this approach are rules for processing information. A
second new element concerns the statistical interpretation of the wave
function. The magnitude of the wave function is interpreted in the usual way.
What is new is that the phase also receives a statistical interpretation: it
is directly related to the entropy of the hidden variables.

Our goal in this paper is to focus on a third new aspect. We discuss how a
dynamics driven by entropy naturally leads to an \textquotedblleft
entropic\textquotedblright\ notion of time. Entropic time is introduced as a
convenient book-keeping device to keep track of the accumulation of change.
Our task here is to develop a model that includes (a) something one might
identify as an \textquotedblleft instant\textquotedblright, (b) a sense in
which these instants can be \textquotedblleft ordered\textquotedblright, (c) a
convenient concept of \textquotedblleft duration\textquotedblright\ measuring
the separation between instants. The welcome new feature is that entropic time
is intrinsically directional. Thus, an arrow of time is generated
automatically. We also discuss the relation between entropic time, which is a
purely inferential device, and the presumably more objective notion of
\textquotedblleft physical\textquotedblright\ time. We argue that for the
pragmatic purpose of predicting the empirically observable correlations among
particles nothing more \textquotedblleft physical\textquotedblright\ than
entropic time is needed.

\section{Entropic dynamics}

The objective is to make inferences about the positions $x\in\mathcal{X}_{N} $
of $N$ particles on the basis of information about some hidden variables
$y\in\mathcal{Y}$ \cite{Caticha 10}. For a single particle the configuration
space $\mathcal{X}$ is Euclidean and 3-dimensional with metric $\gamma
_{ab}=\delta_{ab}/\sigma^{2}$. We will focus on a single particle because the
generalization to $N$ particles is straightforward -- the anisotropy of
$\mathcal{X}_{N}$ is easily represented by including different scale factors
$\sigma_{i}^{2}$ for different particles.

The number, the nature, and the values $y$ of the hidden variables need not be
specified -- they will remain \textquotedblleft hidden\textquotedblright%
\ throughout the analysis. We only need to assume that the unknown $y$ are
described by a probability distribution $p(y|x)$ that depends on the position
$x$ of the particle. To each $x\in\mathcal{X}$ there corresponds a $p(y|x)$
and therefore the set $\mathcal{M=}\{p(y|x);x\in\mathcal{X\}}$ is a
3-dimensional statistical manifold. Most features of $p(y|x)$ turn out to be
irrelevant to the dynamics of $x$; what turns out to be relevant is the
entropy of the hidden variables,
\begin{equation}
S[p,q]=-\int dy\,p(y|x)\log\frac{p(y|x)}{q(y)}=S(x)~,\label{entropy a}%
\end{equation}
where $q(y)$ is some underlying measure which need not be specified further.
Note that $x$ enters as a parameter in $p(y|x)$ so that the entropy is a
function of $x$: $S[p,q]=S(x)$. As we shall see, $S(x)$ will later be
determined from energy considerations so that it is not necessary to know
$p(y|x)$ explicitly.

To find the probability $P(x^{\prime}|x)$ of a short step from $x$ to a nearby
point $x^{\prime}$ we use the method of maximum entropy \cite{Caticha 08}.

\noindent\textbf{The relevant space is }$\mathcal{X\times Y}$. This is because
we want to infer $x^{\prime}$ on the basis of information about the new
$y^{\prime}$. What we need is the joint distribution $P(x^{\prime},y^{\prime
}|x)$ and the appropriate (relative) entropy is
\begin{equation}
\mathcal{S}[P,Q]=-\int dx^{\prime}dy^{\prime}\,P(x^{\prime},y^{\prime}%
|x)\log\frac{P(x^{\prime},y^{\prime}|x)}{Q(x^{\prime},y^{\prime}%
|x)}~.\label{Sppi}%
\end{equation}
The relevant information is introduced through the prior $Q(x^{\prime
},y^{\prime}|x)$ and through suitable constraints on the acceptable posteriors
$P(x^{\prime},y^{\prime}|x)$.

\noindent\textbf{The prior} ref{}lects a state of extreme ignorance: the new
$(x^{\prime},y^{\prime})$ could in principle be anywhere in $\mathcal{X\times
Y}$, and we have no idea how they are related to each other -- knowing
$x^{\prime}$ tells us nothing about $y^{\prime}$ and vice versa. This is
described by a uniform distribution (probabilities are proportional to
volumes). The chosen prior is $Q(x^{\prime},y^{\prime}|x)\propto q(y^{\prime
})$.

\noindent\textbf{The first constraint} on the posterior $P(x^{\prime
},y^{\prime}|x)=P(x^{\prime}|x)P(y^{\prime}|x^{\prime},x)$ establishes the
known relation between $y^{\prime}$ and $x^{\prime}$. It says that
$P(y^{\prime}|x^{\prime},x)=p(y^{\prime}|x^{\prime})\in\mathcal{M}$. The
uncertainty in $y^{\prime}$ depends only on the current position $x^{\prime}$,
and not on any previous value $x$.

\noindent\textbf{The second constraint} ref{}lects the physical fact that
motion is continuous. Motion over large distances happens but only through the
successive accumulation of many short steps. Let $x^{\prime}=x+\Delta x$. We
require that the expectation $\left\langle \Delta\ell^{2}\right\rangle
=\left\langle \gamma_{ab}\Delta x^{a}\Delta x^{b}\right\rangle =\kappa\,$be
some small but for now unspecified numerical value $\kappa$, which we take to
be independent of $x$ in order to ref{}lect the translational symmetry of the
space $\mathcal{X}$.

Varying $P(x^{\prime}|x)$ to maximize $\mathcal{S}[P,Q]$ in (\ref{Sppi})
subject to the two constraints plus normalization gives
\begin{equation}
P(x^{\prime}|x)=\frac{1}{\zeta}e^{S(x^{\prime})-\frac{1}{2}\alpha\,\gamma
_{ab}\Delta x^{a}\Delta x^{b}}~,\label{Prob xp/x}%
\end{equation}
where $\zeta$ is a normalization constant, and the Lagrange multiplier
$\alpha$ is determined in the standard way, $\partial\log\zeta/\partial
\alpha=-\kappa/2$.

$P(x^{\prime}|x)$ shows that in the limit of large $\alpha$ we expect short
steps in essentially random directions with a small anisotropic bias along the
entropy gradient. Expanding the exponent in $P(x^{\prime}|x)$ about its
maximum gives a Gaussian distribution,
\begin{equation}
P(x^{\prime}|x)\propto\exp\left[  -\frac{\alpha}{2\sigma^{2}}\delta
_{ab}\left(  \Delta x^{a}-\Delta\bar{x}^{a}\right)  \left(  \Delta
x^{b}-\Delta\bar{x}^{b}\right)  \right]  ,\label{Prob xp/x b}%
\end{equation}
The displacement $\Delta x^{a}=\Delta\bar{x}^{a}+\Delta w^{a}$ can be
expressed as the expected drift plus a f{}luctuation%
\begin{equation}
\Delta\bar{x}^{a}=\frac{\sigma^{2}}{\alpha}\delta^{ab}\partial_{b}%
S(x)~,\label{ED drift}%
\end{equation}%
\begin{equation}
\left\langle \Delta w^{a}\right\rangle =0\quad\text{and}\quad\left\langle
\Delta w^{a}\Delta w^{b}\right\rangle =\frac{\sigma^{2}}{\alpha}\delta
^{ab}~.\label{ED fluctuations}%
\end{equation}
As $\alpha\rightarrow\infty$ the f{}luctuations become dominant: the drift
$\Delta\bar{x}\propto\alpha^{-1}$ while $\Delta w$ is of order $\alpha^{-1/2}%
$. This implies that, as in Brownian motion, the trajectory is continuous but
not differentiable.

\section{Entropic time}

The necessity to keep track of the accumulation of small changes requires us
to develop appropriate book-keeping tools. Here we show how a dynamics driven
by entropy naturally leads to an \textquotedblleft entropic\textquotedblright%
\ notion of time.

\noindent\textbf{An ordered sequence of instants}

The foundation of all notions of time is dynamics. In entropic dynamics, at
least for infinitesimally short steps, change is given by the transition
probability $P(x^{\prime}|x)$ in eq.(\ref{Prob xp/x b}). The $n$th step takes
us from $x=x_{n-1}$ to $x^{\prime}=x_{n}$. Using the product rule for the
joint probability, $P(x_{n},x_{n-1})=P(x_{n}|x_{n-1})P(x_{n-1})$, and
integrating over $x_{n-1}$, we get
\begin{equation}
P(x_{n})=%
{\textstyle\int}
d^{3}x_{n-1}\,P(x_{n}|x_{n-1})P(x_{n-1})~.\label{CK a}%
\end{equation}
This equation is a direct consequence of the laws of probability. However, if
$P(x_{n-1})$ happens to be the probability of different values of $x_{n-1} $
\emph{at a given instant of entropic time }$t$, then we will interpret
$P(x_{n})$ as the probability of values of $x_{n}$ at the \textquotedblleft
later\textquotedblright\ instant of entropic time $t^{\prime}=t+\Delta t$.
Accordingly, we write $P(x_{n-1})=\rho(x,t)$ and $P(x_{n})=\rho(x^{\prime
},t^{\prime})$ so that
\begin{equation}
\rho(x^{\prime},t^{\prime})=%
{\textstyle\int}
dx\,P(x^{\prime}|x)\rho(x,t)\label{CK b}%
\end{equation}
Nothing in the laws of probability that led to eq.(\ref{CK a}) forces this
interpretation on us---this is an independent assumption about what
constitutes time in our model. We use eq.(\ref{CK b}) to define what we mean
by an instant:\emph{\ if the distribution }$\rho(x,t)$\emph{\ refers to one
instant, then the distribution }$\rho(x^{\prime},t^{\prime})$\emph{\ defines
what we mean by the \textquotedblleft next\textquotedblright\ instant }and
eq.(\ref{CK b}) allows entropic time to be constructed one instant after
another.\emph{\ }

\noindent\textbf{The arrow of entropic time}

Time constructed according to eq.(\ref{CK b}) is remarkable in yet another
respect: the inference implied by $P(x^{\prime}|x)$ in eq.(\ref{Prob xp/x b})
incorporates an intrinsic directionality in entropic time: there is an
absolute sense in which $\rho(x,t)$\ is prior and $\rho(x^{\prime},t^{\prime
})$\ is posterior.

Suppose we wanted to find a time-reversed evolution. We would write
\begin{equation}
\rho(x,t)=%
{\textstyle\int}
dx^{\prime}\,P(x|x^{\prime})\rho(x^{\prime},t^{\prime})\,.
\end{equation}
This is perfectly legitimate but in order to be correct $P(x|x^{\prime})$
cannot be obtained from eq.(\ref{Prob xp/x b}) by merely exchanging $x$ and
$x^{\prime}$. According to the rules of probability theory $P(x|x^{\prime})$
is related to eq.(\ref{Prob xp/x b}) by Bayes' theorem,
\begin{equation}
P(x|x^{\prime})=\frac{P(x)}{P(x^{\prime})}P(x^{\prime}|x)~.
\end{equation}
In other words, one of the two transition probabilities, either $P(x|x^{\prime
})$ or $P(x|x^{\prime})$, \emph{but not both}, can be given by the maximum
entropy distribution eq.(\ref{Prob xp/x b}). The other is related to it by
Bayes' theorem. I hesitate to say that this is what breaks the time-reversal
symmetry because the symmetry was never there in the first place. There is no
symmetry between prior and posterior; there is no symmetry between the
inferential past and the inferential future.

The puzzle of the arrow of time has a long history (see \emph{e.g.}
\cite{Price96 Zeh01}). The standard question is how can an arrow of time be
derived from underlying laws of nature that are symmetric? Entropic dynamics
offers a new perspective because it does not assume any underlying laws of
nature -- whether they be symmetric or not -- and its goal is not to explain
the asymmetry between past and future. The asymmetry is the inevitable
consequence of entropic inference. From the point of view of entropic dynamics
the challenge does not consist in explaining the arrow of time, but rather in
explaining how it comes about that despite the arrow of time some laws of
physics turn out to be reversible. Indeed, even when the derived laws of
physics -- in our case, the Schr\"{o}dinger equation -- turns out to be fully
time-reversible, \emph{entropic time itself only f{}lows forward}.

\noindent\textbf{A convenient scale of time}

Duration is defined so that motion looks simple. Since longer steps presumably
take a longer time, specifying the interval $\Delta t$ between successive
instants amounts to specifying the multiplier $\alpha(x,t)$ in terms of
$\Delta t$. For large $\alpha$ the dynamics is dominated by the f{}luctuations
$\Delta w$. In order that the f{}luctuations $\left\langle \Delta w^{a}\Delta
w^{b}\right\rangle $ ref{}lect the symmetry of translations in space and time
we choose $\alpha$ independent of $x$ and $t$, $\alpha(x,t)=\tau/\Delta
t=\operatorname{constant}$, where $\tau$ is a constant that fixes the units of
time. With this choice the dynamics is indeed simple: $P(x^{\prime}|x)$
in\ (\ref{Prob xp/x b}) becomes a standard Wiener process. The displacement is
$\Delta x=b(x)\Delta t+\Delta w$, where $b(x)$ is the drift velocity,
\begin{equation}
\langle\Delta x^{a}\rangle=b^{a}\Delta t\quad\text{with}\quad b^{a}%
(x)=\frac{\sigma^{2}}{\tau}\,\delta^{ab}\partial_{b}%
S(x)~,\label{drift velocity}%
\end{equation}
and $\Delta w$ is the f{}luctuation,
\begin{equation}
\left\langle \Delta w^{a}\right\rangle =0\quad\text{and}\quad\left\langle
\Delta w^{a}\Delta w^{b}\right\rangle =\frac{\sigma^{2}}{\tau}\Delta
t\,\delta^{ab}~.\label{fluc}%
\end{equation}

The formal similarity to Nelson's stochastic mechanics \cite{Nelson
66}-\cite{Nelson 85} is evident. The new elements here are (1) that
eq.(\ref{drift velocity}) has been derived rather than postulated, and (2)
that the previously uninterpreted scalar function $S(x)$ now receives a
definite interpretation as the entropy of the hidden variables.

\section{The Schr\"{o}dinger equation}

The evolution of $\rho(x,t)$ obtained by iterating (\ref{CK b}) together with
(\ref{drift velocity})-(\ref{fluc}) is given by the Fokker-Planck equation
(FP) which can be written as a continuity equation, \cite{Chandrasekhar 43}%
\begin{equation}
\partial_{t}\rho=-\partial_{a}\left(  v^{a}\rho\right)  ~,\label{FP}%
\end{equation}
where $v^{a}=b^{a}+u^{a}$, the \emph{current velocity}, is given in terms of
the drift velocity $b^{a}$, eq.(\ref{drift velocity}), and the \emph{osmotic
velocity}
\begin{equation}
u^{a}\overset{\text{def}}{=}-\frac{\sigma^{2}}{\tau}\partial^{a}\log\rho
^{1/2}~.\label{osmo}%
\end{equation}
Thus the probability f{}low $\rho v^{a}$ has two components, one is the drift
current $\rho b^{a}$, and the other is the diffusion current, $\rho
u^{a}=-\frac{\sigma^{2}}{2\tau}\partial^{a}\rho$. Using (\ref{drift velocity})
and (\ref{osmo}) the current velocity can be expressed as a gradient too,
\begin{equation}
v^{a}=\frac{\sigma^{2}}{\tau}\,\partial^{a}\phi\quad\text{where}\quad
\phi=S-\log\rho^{1/2}\label{v and phase}%
\end{equation}

As long as the geometry of the statistical manifold $\mathcal{M}$ is rigidly
fixed the density $\rho(x,t)$ is the only degree of freedom available and the
dynamics described by the FP equation (\ref{FP}) is just standard diffusion --
quantum mechanics requires a second degree of freedom.

The natural solution is to allow the manifold $\mathcal{M}$ to participate in
the dynamics. Then the entropy of the hidden variables becomes a second
dynamical field, $S=$ $S(x,t)$. Quantum dynamics consists of the coupled
evolution of $\rho(x,t)$ and $S(x,t)$. An equivalent but more convenient
choice of variables is the scalar function $\phi=S-\log\rho^{1/2}$ in
(\ref{v and phase}).

To specify the dynamics of the manifold $\mathcal{M}$ we follow Nelson
\cite{Nelson 79} and impose that the dynamics be \textquotedblleft
conservative,\textquotedblright\ that is, one requires the conservation of a
certain functional $E[\rho,\phi]$ of $\rho(x,t)$ and $\phi(x,t)$ that we will
call the \textquotedblleft energy\textquotedblright. In the non-relativistic
regime the \textquotedblleft energy\textquotedblright\ functional is
\cite{Caticha 10}%
\begin{equation}
E[\rho,\phi]=\int d^{3}x\,\rho(x,t)\left(  \frac{1}{2}mv^{2}+\frac{1}{2}%
mu^{2}+V(x)\right)  ~,\label{energy b}%
\end{equation}
which includes terms that are at time reversal invariant and at most quadratic
in the velocities $v^{2}$ and $u^{2}$, and where $V(x)$ represents a
\textquotedblleft potential\textquotedblright\ energy. The coefficient $m$ is
related to other constants in the theory by $m=\eta\tau/\sigma^{2}$, where
$\sigma^{2}$ is the length scale in the metric of $\mathcal{X}$ (required for
the squared velocities), and the new constant $\eta$ is introduced to take
care of units -- if $E$ is given in units of energy, then $m$ has units of
mass. It is possible to assign different coefficients to the $v^{2}$ and
$u^{2}$ terms but this does not change the final conclusions. (For further
details see \cite{Caticha 10}.) In these units the current and osmotic
velocities, eqs.(\ref{v and phase}) and (\ref{osmo}), are
\begin{equation}
mv_{a}=\eta\,\partial_{a}\phi\quad\text{and\quad}mu_{a}=-\eta\partial_{a}%
\log\rho^{1/2}~.\label{curr osmo}%
\end{equation}
The FP equation and the energy $E$ become
\begin{equation}
\dot{\rho}=-\partial_{a}\left(  \rho v^{a}\right)  =-\frac{\eta}{m}%
\partial^{a}\left(  \rho\partial_{a}\phi\right)  ~\label{SEa}%
\end{equation}
and%
\begin{equation}
E=\int dx\,\rho\left(  \frac{\eta^{2}}{2m}(\partial_{a}\phi)^{2}+\frac{\mu
\eta^{2}}{2m^{2}}(\partial_{a}\log\rho^{1/2})^{2}+V\right)  ~.\label{energy c}%
\end{equation}
Imposing that $\dot{E}=0$ for arbitrary choices of $\dot{\rho}$ leads after
some algebra \cite{Caticha 10} to
\begin{equation}
\eta\dot{\phi}+\frac{\eta^{2}}{2m}(\partial_{a}\phi)^{2}+V-\frac{\eta^{2}}%
{2m}\frac{\nabla^{2}\rho^{1/2}}{\rho^{1/2}}=0~.\label{SEb}%
\end{equation}

Equations (\ref{SEa}) and (\ref{SEb}) are the coupled dynamical equations we
seek. The evolution of $\rho(x,t)$, eq.(\ref{SEa}), is guided by $\phi(x,t) $;
and the evolution of $\phi(x,t)$, eq.(\ref{SEb}), is determined by $\rho(x,t)$.

The two real equations can be condensed into a single complex equation by
combining $\rho$ and $\phi$ into a complex function $\Psi=\rho^{1/2}\exp
(i\phi)$. Computing the time derivative $\dot{\Psi}$ and using eqs.
(\ref{SEa}) and (\ref{SEb}) leads to the Schr\"{o}dinger equation,
\begin{equation}
i\eta\frac{\partial\Psi}{\partial t}=-\frac{\eta^{2}}{2m}\nabla^{2}\Psi
+V\Psi~,\label{SE}%
\end{equation}
which allows us to identify $m$ with the particle mass and $\eta$ with
Planck's constant, $\eta=\hbar$. Therefore the Schr\"{o}dinger equation
describes a conservative diffusion driven by the entropy of the hidden variables.

Other attempts to derive quantum theory start from an underlying, perhaps
stochastic, classical mechanics. The entropic dynamics approach is different
in that it does not assume an underlying classical substrate; entropic
dynamics provides a derivation of \emph{both Schr\"{o}dinger's equation and
also Newton's }$F=ma$\emph{.} Classical mechanics is recovered in the usual
limits of $\hbar\rightarrow0$ or $m\rightarrow\infty$. Indeed, writing
$S_{HJ}=\eta\phi$ in eq.(\ref{SEb}) and letting $m\rightarrow\infty$ with
$S_{HJ}/m$ fixed leads to the classical Hamilton-Jacobi equation%
\begin{equation}
\dot{S}_{HJ}+\frac{1}{2m}(\partial_{a}S_{HJ})^{2}+V=0~,\label{HJ}%
\end{equation}
while eq.(\ref{curr osmo}) gives $mv=\,\partial S_{HJ}$ and $u=0$. Since
$mv\approx mb=\eta\partial S$ we see that \emph{classical particles}
\emph{move along the gradient of the entropy of the hidden variables}, with
vanishing f{}luctuations $\left\langle \Delta w^{a}\Delta w^{b}\right\rangle
=\frac{\eta}{m}\Delta t\,\delta^{ab}\rightarrow0$. The classical
Hamilton-Jacobi action is interpreted as the entropy of the hidden variables,
$S_{HJ}=\eta S$.

\section{Entropic time \emph{vs.} \textquotedblleft physical\textquotedblright%
\ time}

We can now return to the question of the actual connection between entropic
and physical time. If the order of the sequence of inferential steps generated
by the Fokker-Planck equation is not the same as the order relative to a
presumably more fundamental \textquotedblleft physical\textquotedblright%
\ time, why should `entropic time' deserve to be called `time' at all?

The answer is that we are typically concerned with systems that include, in
addition to the particles of interest, also at least one other system that one
might call a \textquotedblleft clock\textquotedblright. The goal is to make
inferences about correlations between the particles and the various states of
the clock. Whether the inferred sequence of states of the particle-clock
composite agrees or not with an order in \textquotedblleft
physical\textquotedblright\ time turns out to be quite irrelevant. It is only
the correlations among the particles and the clock that are observable and not
their \textquotedblleft absolute\textquotedblright\ order.

Earlier we introduced a concept of simultaneity through the density
$\rho(x,t)$ which gives the probability of different values of $x$ at the same
instant. We can now tackle the same issue from a different angle. Since the
particle can follow different paths we need a criterion to decide whether an
event $x^{\prime}$ reached along one path is earlier or later than another
event $x^{\prime\prime}$ reached along a different path. This is where a clock
comes in handy. The clock could be, for example, just a sufficiently massive
particle that follows a deterministic trajectory, $x_{C}=\bar{x}_{C}(t)$, and
remains largely unaffected by the motion of other particles.

The idea is that when we compute the probability that, say, after $n$ steps
the particle is found at the point $x_{n}$ we implicitly \emph{assume} that
its three coordinates $x_{n}^{1}$, $x_{n}^{2}$, and $x_{n}^{3}$ are attained
\emph{simultaneously}. We make the same \emph{assumption} in the larger
configuration space of the composite particle-clock system, $x_{n}^{A}%
=(x_{n}^{a},x_{Cn}^{\alpha})$. The particle coordinates $x_{n}^{a}$
($a=1,2,3$) are assumed to be simultaneous with the clock coordinates
$x_{Cn}^{\alpha}$ ($\alpha=4,5,\ldots$). Thus, when we say that at the $n$th
step the particle is at $x_{n}^{a}$ while the clock is at $x_{Cn}^{\alpha}$ it
is implicit that these positions are attained \emph{at the same time}.

By \textquotedblleft the time is $t$\textquotedblright\ we just mean that
\textquotedblleft the clock is in its state $x_{C}=\bar{x}_{C}(t)$%
.\textquotedblright\ We say that the possible event that the particle reached
$x^{\prime}$ along one path is simultaneous with another possible event
$x^{\prime\prime}$ reached along a different path when both are simultaneous
with the same state $\bar{x}_{C}(t)$ of the clock. We usually omit referring
to the clock and just say that $x^{\prime}$ and $x^{\prime\prime}$ happen
\textquotedblleft at the same time $t$.\textquotedblright\ This justifies
using the distribution $\rho(x,t)$ as the definition of an instant of time.

In the end the justification for the assumptions underlying entropic dynamics
lies in their empirical success. The ordering scheme provided by entropic time
allows one to predict correlations. Since these predictions, which are given
by the Schr\"{o}dinger equation, turn out to be tremendously successful one
concludes that nothing deeper or more \textquotedblleft
physical\textquotedblright\ than entropic time is needed.

\section{Some remarks}

Since it is the correlations between particles and clocks that are empirically
accessible and not their absolute order we argued that entropic time is all we
need. The time $t$ that appears in the laws of physics, be it the
Schr\"{o}dinger equation or Newton's $F=ma$, is entropic time.

Julian Barbour in his relational approach to time in the context of classical
dynamics \cite{Barbour 94} has made the strong claim that time is not real,
that time is a mere illusion. Barbour's claim is much stronger than ours. The
reason is that from Barbour's perspective the laws of physics directly
ref{}lect the laws of nature. Then the absence of physical time in the laws of
physics is strong evidence for its non-existence. In this view, just as the
ether of an earlier generation, time is an obsolete feature of outdated physics.

From the perspective of entropic dynamics, however, the laws of physics are
not laws of nature but mere rules of inference. The fact that it is entropic
time that appears in the laws of physics does not entitle us to conclude that
physical time does not exist. We can only conclude that physical time is not
the carrier of any piece of information that happens to be relevant for our
current inferences.

One possibility is that physical time is an illusion, which explains why it
would be irrelevant in the first place. Another possibility is that physical
time actually exists and is identical with entropic time -- which explains why
it shows up in the equations of physics. Whether physical time is real or a
mere illusion remains a question for the future.

\noindent\textbf{Acknowledgments:} I would like to thank Carlos Rodr\'{\i}guez
and N\'{e}stor Caticha for their many insights into entropy and inference. My
deep appreciation also goes to C. Cafaro, A. Giffin, P. Goyal, D. Johnson, K.
Knuth, S. Nawaz, and C.-Y. Tseng for many discussions on the ideas that
eventually evolved into this paper.

\end{document}